\documentclass[showpacs,aps,prb,twocolumn]{revtex4}
\usepackage{epsfig}

\begin{document}

\title{Epitaxially strained [001]-(PbTiO$_3$)$_1$(PbZrO$_3$)$_1$ superlattice and PbTiO$_3$
from first principles}

\author{Claudia Bungaro}
\email{bungaro@physics.rutgers.edu}
\author{K. M.~Rabe}
\affiliation{Department of Physics and Astronomy, Rutgers University, 
Piscataway, NJ 08854-8019, USA.}

\begin{abstract}
The effect of layer-by-layer heterostructuring and epitaxial strain
on lattice instabilities and related ferroelectric properties is 
investigated from first principles for the 
[001]-(PbTiO$_3$)$_1$(PbZrO$_3$)$_1$ superlattice
and pure PbTiO$_3$ on a cubic substrate.
The results for the superlattice
show an enhancement of the stability of the monoclinic
r-phase with respect to pure PbTiO$_3$.
Analysis of the lattice instabilities of the
relaxed centrosymmetric reference structure 
computed within density functional perturbation
theory suggests
that this results from the presence of two
unstable zone-center modes, one confined
in the PbTiO$_3$ layer and one in the PbZrO$_3$ layer, which
produce in-plane and normal components of the
polarization, respectively.
The zero-temperature dielectric response is computed
and shown to be enhanced not only near the phase
boundaries, but throughout the r-phase.
Analysis of the analogous calculation for
pure PbTiO$_3$ is consistent with this interpretation,
and suggests useful approaches to engineering 
the dielectric properties of artificially structured
perovskite oxides.

\end{abstract}
\date{\today}
\pacs{  }

\maketitle

%
%
%

\section{INTRODUCTION}
Perovskite oxide superlattices
are expected to have properties distinct from
those of the bulk constituents. Due to lattice mismatch, constituent layers
below the critical thickness generally are in highly strained states. As
can be confirmed by first-principles
studies of bulk materials, the lattice instabilities and properties
of bulk perovskites are extremely sensitive to the strain state. The degree 
of strain can be further tuned by varying the lattice constant of the
substrate on which the superlattice is grown.

This discussion is of practical interest as it has been 
shown that perovskite oxide superlattices can indeed be grown
with atomic scale precision \cite{Schlom,Eckstein,Christen},
opening up a highly varied family of artificially structured
materials for investigation. 
Thus, first-principles prediction of superlattice properties 
can directly lead to productive
interactions with experimentalists and progress in materials
design\cite{BT,BTST}.

In this work, we consider the simplest possible (and as yet hypothetical) 
[001]-(PbTiO$_3$)$_m$(PbZrO$_3$)$_n$ superlattice with $m$=1 and $n$=1.
%
\begin{figure}
\centerline{\epsfig{figure=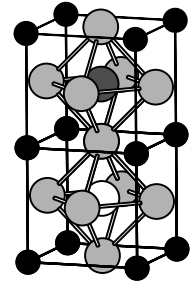,width=4.5cm, angle=-90}}
\caption{The ideal double-perovskite 10-atom unit cell of the
[001]-(PbTiO$_3$)$_1$(PbZrO$_3$)$_1$ superlattice. The lead, oxygen, titanium
and zirconium atoms are indicated by small solid circles, light grey circles,
dark grey circles and open circles, respectively. The oxygen octahedra and
the ideal perovskite 5-atom cells are outlined. The pseudoperovskite $c/a$ is defined
as half the unit cell $c/a$. Note that this is the same
structure considered as a ordered realization of PZT in 
Ref.~\protect\onlinecite{KrakauerWu}.
}
\label{fig_supercell}
\end{figure}
As shown in Fig.~\ref{fig_supercell}, 
this consists of single unit cell layers of PbZrO$_3$ (PZ) and 
PbTiO$_3$ (PT) alternating 
along [001]. The equilibrium cubic lattice constant of bulk PZ is significantly
greater than that of PT, with the computed values being $a_{0,PZ}$=7.77 a.u. and
$a_{0,PT}$=7.37 a.u., leading to a 5\% mismatch 
(the corresponding experimental values are 7.81 a.u., 7.50 a.u, and 4\%).
At their equilibrium lattice constants, both cubic PT and cubic PZ have an unstable
three-fold degenerate polar phonon mode at the zone center.
The lattice mismatch means that in the superlattice,  PZ is under compressive in-plane stress, 
while PT is under tensile in-plane stress.
If the structure is allowed to relax without breaking the central mirror
plane symmetry (i.e. in space group P4/mmm), 
the in-plane lattice constant takes an intermediate value of 7.57 a.u..
The PZ unit cell elongates,
giving a local tetragonal unit cell with $c/a$ $>$ 1, and conversely,
the PT unit cell will shorten, resulting in a local tetragonal unit cell with $c/a$ $<$ 1.
This lowering of unit cell symmetry leads to a splitting of the three-fold degenerate unstable mode in each layer,
with the lower frequency mode(s) along the long direction(s).
Thus, on qualitative
grounds, one expects at least two unstable $\Gamma$ modes of different symmetry, one associated
with polarization along the normal in the PZ layer, and a second,
two-fold degenerate, mode producing polarization in the plane of the
PT layer. DFPT calculations of the phonon dispersion of the 
relaxed centrosymmetric superlattice show two unstable zone-center polar modes, one
with in-plane polarization confined in the PT layer and the other with polarization
along the normal confined in the PZ layer, quantitatively confirming this picture \cite{Bungaro}.
The energy of the system can be lowered by either of these
modes alone; the actual ground state might involve choosing one, or the
other, or coupling both together.

First principles ground-state-structure calculations have previously been performed for this 
system, regarding the 1:1 [001] ordered PZT 50/50 supercell as an approximate realization of 
the PZT solid solution \cite{KrakauerWu}. Under the constraint of experimental volume,
$c/a$ = 1.035 and $\alpha=\beta=$ 90 degrees, it was found that the relaxed structure 
has monoclinic $Cm$ symmetry, with internal coordinates
in reasonable agreement with those experimentally determined for the monoclinic PZT phase \cite{Noheda}. 
Connecting this result to our discussion in terms of lattice instabilities, we see that
the ground state is in fact obtained by freezing in the two modes together, 
giving a total polarization ${\bf P} = (P,P,P_z)$ rotated away from the c-axis.
Piezoelectric coefficients computed for this structure in Ref. \onlinecite{KrakauerWu} have very large values.
This is attributed to the low energy for polarization rotation; this mechanism was first
suggested to explain the high piezoelectric response of single crystal PMN-PT \cite{FuCohen,Shrout}
and accounts for the enhancement not only of the piezoelectric
response in the monoclinic phase of the superlattice, but also, as we will show below, for
the enhancement of the dielectric response.

Under an epitaxial constraint imposed by a cubic [001] substrate, 
the in-plane lattice vectors are equal in length and make an angle
of 90 degrees, while
the system is otherwise free to relax. 
Our qualitative picture of the superlattice instabilities suggests that the
two modes should respond differently to changes in the substrate lattice constant.
As the substrate lattice constant increases, the mode with polarization
along the normal should become higher in frequency, while the mode with
in-plane polarization should soften. 
This change in relative stability is expected to affect the direction
of the polarization in the ground state structure,
in a manner consistent with previous calculations in which a different structural
constraint, on $c/a$, was imposed.\cite{KrakauerWu}

In this paper, we present a quantitative first-principles analysis
of the effect of varying in-plane strain on the epitaxially constrained
[001]-(PbTiO$_3$)$_1$(PbZrO$_3$)$_1$ superlattice.
In section II, we describe the details of the computations.
In section III, we present results of the optimized structural parameters for in-plane 
lattice constants $a_\parallel$ ranging 
from well below $a_{0,PT}$, at which the PT unit cell is expected to be approximately
cubic, to above $a_{0,PZ}$. This wide range of strain allows us more fully to 
investigate the effects of strain on structure and properties.
As we will see, there is a sequence of phases, from the tetragonal c-phase at the smallest
values of $a_\parallel$, through a monoclinic r-phase with increasing $a_0$ and finally
to an orthorhombic aa-phase.
The locations of the phase boundaries are determined by computing the Hessian matrix
of each optimized structure, comprised of the zone-center force-constant matrix, 
the elastic constants and the quadratic-order coupling between zone-center atomic 
displacements and homogeneous strain, and extrapolating to zero eigenvalue. 
Near the phase boundaries and throughout the r-phase
the dielectric and piezoelectric responses are high, the sensitivity
to applied fields and stresses resulting from the ease with which the polarization
can be changed through rotation.
We relate these results to the lattice instabilities of the centrosymmetric reference
structure obtained from DFPT computation of the ${\bf q} = 0$ frequencies.
The behavior of the superlattice is compared  
with that of pure PT, which is seen to exhibit a qualitatively different behavior.
In section IV, we discuss this distinction using a two-mode vs one-mode
picture, and suggest some ways that these results can be used in materials
design.

\section{Method}

To predict the ground state structure and zone-center phonon frequencies, we use 
density functional theory (DFT) and density functional perturbation theory (DFPT)\cite{hks,Dfptrev}
within the local density approximation (LDA).\cite{CAPZ} 
Calculations in this work have been done using the PWscf package.~\cite{PWSCF}
We use Vanderbilt ultrasoft pseudopotentials \cite{ultrasoft}
to describe the interaction between ionic cores and valence electrons, and
plane-wave basis set with kinetic energy cut-off of 35 Ry.
The augmentation charges, required by the use of ultrasoft pseudopotentials, 
are expanded with an energy cutoff of 350 Ry. 
Brillouin zone (BZ) integrations are performed with a 6$\times$6$\times$3 
Monkhorst-Pack mesh.

The polarization {\bf P} is computed using the linearized expression
{\bf P$_\alpha$}= $\sum_{i\beta}$Z$_{i,\alpha\beta}^*$ {u}$_{i\beta}$, where {\bf u}$_i$ are 
the atomic displacements of the 
optimized structure relative to the centrosymmetric reference structure (described below),
and Z$^*$ are the atomic Born effective charge tensors.
The values of polarization reported below are computed with the averaged
effective charge tensors of pure PT and pure PZ given in Table II of Ref.~\onlinecite{Bungaro},
which have been shown to be a good approximation to those of the superlattice.

The epitaxial constraint from a cubic substrate is treated implicitly by 
constraining the in-plane lattice constant of the system.
We consider a range of in-plane strains from -6\% to +4\%. 
As the reference in-plane lattice parameter to define the in-plane strain,
we use the equilibrium in-plane lattice parameter computed
for the optimized polar structure, $a_0$ = 7.73 a.u., for the superlattice 
and $a_0$ = 7.32 a.u. for PT. 
Thus, the system at zero in-plane strain has the minimum elastic energy.
 
For each value of the in-plane strain, we perform the structural optimization 
of the superlattice and of pure PT in two stages. 
First, the ideal double-perovskite 10-atom unit cell of the superlattice is relaxed without 
any symmetry breaking beyond that associated with the Zr/Ti ordering 
(tetragonal space group $P4/mmm$, $\#$123). Similarly, the perovskite 5-atom unit cell of
PT is relaxed in space group $P4/mmm$.
These will be referred to as the centrosymmetric reference structures.
The zone center phonon modes of these structures are computed using DFPT.
This shows the presence or absence of unstable
modes that provide a guide to energy-lowering distortions.

Then, the symmetry is broken to allow relaxation into a lower symmetry
phase, which we label following the notation of Pertsev et al..\cite{Pertsev}
For the whole range of in-plane strain, we consider relaxations into the
c-phase (tetragonal space group $P4mm$, $\#$99) with polarization $P\hat z$, 
the aa-phase (orthorhombic space group $Cmm2$, $\#$35), 
with polarization $P(\hat x+\hat y)$,
and the r-phase (monoclinic space group $Cm$,$\#$8), with polarization 
$P (\hat x + \hat y )+ P_z \hat z$.
We also investigated the ac-phase (monoclinic space group Pm, $\#$6), with polarization 
$P \hat x + P_z \hat z$,
and the a-phase (orthorhombic space group $Pmm2$, $\#$25), with polarization $P\hat x$; these
two phases are found to be energetically unfavorable and do not appear in the present 
phase diagram.
In principle, an equilibrium c-phase can be obtained by optimizing structural parameters
from a starting r-phase or ac-phase (and an equilibrium aa-phase from a starting r-phase) 
though in practice more accurate results can be obtained by studying each of the
space groups separately.
Compatible k-point sampling allows us directly to compare the
energies of these structures.
Although the angle $\gamma$ between
the c axis and the plane can in principle vary, in the total energy
calculations it is held fixed
at 90 degrees to simplify the calculation. The main effect 
of this approximation is to reduce 
the range of stability of the r-phase as a
function of in-plane strain; for the systems considered here, we describe below how we 
compute the effect of strain
through analysis of the Hessian matrix.

The zone-center dynamical matrices are computed for the optimized structures using DFPT.
The zone-center phonon modes, obtained by diagonalizing the dynamical matrices, 
are used according to the formalism of Ref. \onlinecite{dielectric}
to compute the phonon contribution to the zero-temperature
constant-strain static dielectric tensor and to study its dependence upon epitaxial strain.
We also used these results for a systematic study of the stability
of the various phases as a function of in-plane strain.
The full Hessian matrix of the second derivatives of the energy with respect
to lattice-periodic atomic displacements and homogeneous strain is constructed
with DFPT for the atomic displacements and finite difference stress and force
results for the derivatives involving strain.
Tracking the lowest eigenvalue of the Hessian matrix, we can
accurately identify the instability point of a given phase
as the in-plane strain where the lowest eigenvalue becomes negative \cite{Garcia}.
This allows a more accurate determination of the position of the phase boundary
than total energy calculations alone, which are limited by computational precision, 
and also allows us to include the effects of the coupling to shear strain 
$\eta_4$ and $\eta_5$ in the r-phase,
equivalent to allowing $\gamma$ to relax from 90 degrees.

%
\begin{figure}[h]
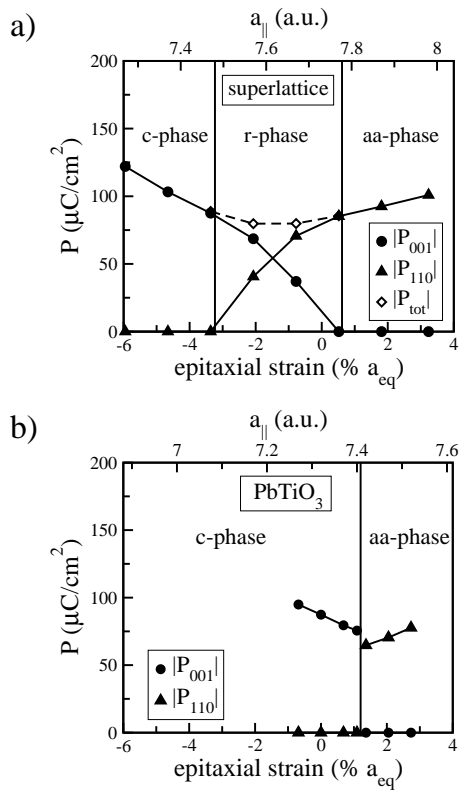

\vbox{
\centerline{\epsfig{figure=Fig2a.eps,width=6cm}}
\vspace{0.2cm}
\centerline{\epsfig{figure=Fig2b.eps,width=6cm}}
}
\caption{Components of the polarization along the [001] and [110] directions, and its 
magnitude $\vert {\bf P}_{\rm tot}\vert$, as a function of in-plane strain for
(a) the [001]-(PbTiO$_3$)$_1$(PbZrO$_3$)$_1$ superlattice, and (b) pure PT. }
\label{fig_gs}
\end{figure}
%
\section{Results}

This section is organized as follows. In A, we present the calculations
of the equilibrium structural parameters and polarization as a function of in-plane strain.
Calculations of zone-center phonon modes, elastic constants, and coupling
between the atomic displacements and homogeneous strain are performed to
carry out the stability analysis and precisely locate the phase boundaries.
In B, we present the results of the zone-center phonons in the centrosymmetric
reference structure as a function of in-plane strain and use them to interpret
the phase diagram. In C, we present the in-plane-strain dependence of the
static dielectric tensor. Results in each subsection are
presented first for the superlattice and then, for comparison, for pure PT.

\subsection{Structural parameters and phase diagram}
%
\begin{figure*}[!]
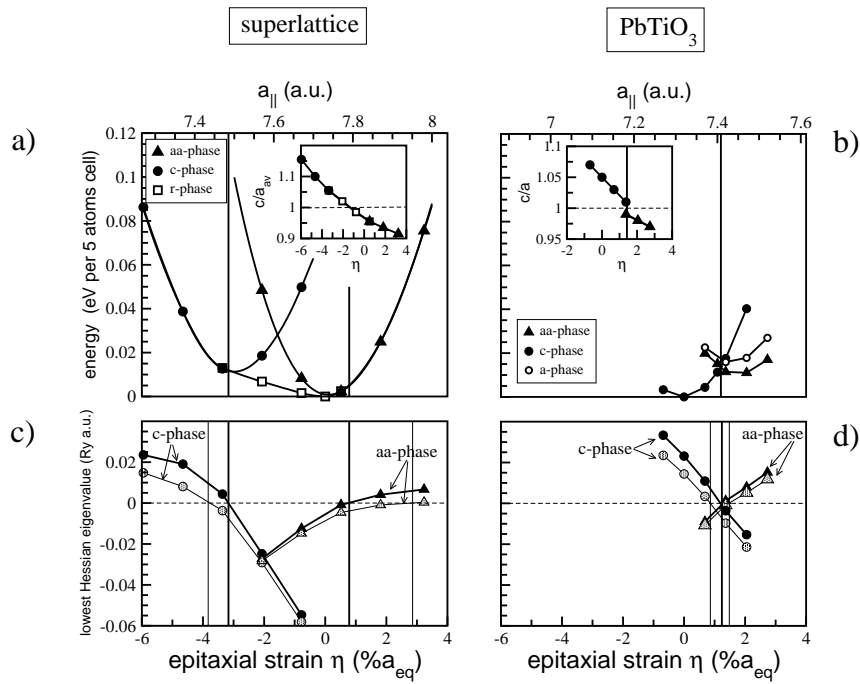

\vbox{
\centerline{
\hbox{
\psfig{figure=Figs3a-c.eps,height=9cm}
\hspace{0.3cm}
\psfig{figure=Figs3b-d.eps,height=9cm}
}
}
}
\caption{The energy as a function of in-plane strain for the c, aa and r phases,
for (a) the [001]-(PbTiO$_3$)$_1$(PbZrO$_3$)$_1$ superlattice and (b) pure PT.
The same scales are used for both systems to facilitate comparison.
The insets
show the tetragonality
$c/a$ as a function of in-plane strain.
The lowest eigenvalue of the Hessian, for (c) the
[001]-(PbTiO$_3$)$_1$(PbZrO$_3$)$_1$ superlattice and (d) pure PT.
The circles are for the c-phase and the triangles for the aa-phase.
The solid symbols are the eigenvalues when only tetragonal strain is allowed,
corresponding to the first-principles results in (a) and (b). The open symbols
show the change when coupling to shear strain is included, resulting
in an enhancement of the stability of the r-phase.
}
\label{fig_energy}
\end{figure*}
In Fig.~\ref{fig_gs}(a) we show the structural phase transitions of the
superlattice with
varying in-plane strain. The symmetry is evident from the nonzero components 
of the polarization. There is a continuous (second-order) transition from 
the c-phase, stable at large compressive in-plane strain, to the r-phase.
In the r-phase, the polarization has almost constant magnitude and
rotates smoothly in the ($1\overline 1 0$) plane from $\hat z$ to $\hat x + \hat y$,
followed by a 
second continuous transition
from the r-phase to the aa-phase, stable at large
tensile in-plane strain. 
The in-plane-strain dependence of the energy in each of these three symmetries is shown in 
Fig.~\ref{fig_energy}(a).
The a-phase and ac-phase energies computed at $a_\parallel$ = 7.57 a.u. were found to be
significantly higher than that of the r-phase and thus are not included in further discussion.
Accurate positions of the second-order c-r and r-aa phase boundaries can be best obtained
not from intersection points
of the total energy curves, but by interpolating
the dependence of the quadratic-order energy around the equilibrium phase at
each in-plane strain to obtain the in-plane strain at which the instability occurs.

We compute the Hessian matrix (phonons plus tetragonal strain, the latter coupling 
being zero for the c-phase and the aa-phase) and plot the lowest eigenvalue
in Fig.~\ref{fig_energy}(c).
The zero-crossing of the eigenvalues can be seen to correspond to the points
where the r-phase becomes more stable in Fig.~\ref{fig_energy}(a); 
in the case of the r-aa transition the critical in-plane strain
is clearly much easier to determine accurately in Fig.~\ref{fig_energy}(c).
To test the effect of fixing $\gamma$ to 90 degrees, we also compute the
full Hessian matrix, including coupling to shear strains $\eta_4$ and $\eta_5$,
and plot the lowest eigenvalue in Fig.~\ref{fig_energy}(c). 
It can be seen that the effect of the additional strain relaxation is to 
expand the region of stability of the r-phase, 
the effect being slight for the c-r 
boundary and more significant for the r-aa boundary.

The corresponding results for pure PbTiO$_3$ present an informative contrast
to those for the superlattice.
In Fig.~\ref{fig_gs}(b) we show the structural phase transitions of pure PT with
varying in-plane strain. The system makes a weakly first-order direct transition from the c-phase at 
large compressive in-plane strain to the aa-phase at large
tensile in-plane strain. The intermediate r-phase present in the superlattice has
been eliminated.
There are small discontinuities in the $c/a$ ratio (inset of Fig.~\ref{fig_energy}(b))
and the magnitude of the
polarization at the phase transition. 
The optimized energies for each symmetry as a function of in-plane
strain are shown in Fig.~\ref{fig_energy}(b). Optimization in the monoclinic space
group corresponding to the r-phase produces either the c-phase or the aa-phase 
depending on in-plane strain; these energies lie very slightly higher than those obtained
by optimizing in the tetragonal space groups, reflecting the convergence criterion for
the relaxation.
We can confirm the weakly discontinuous nature of this transition by computing the
Hessian matrix (phonons plus tetragonal strain) and plotting the eigenvalues
in Fig.~\ref{fig_energy}(d). At the transition in-plane strain, each of the two
phases is marginally stable. 
To test the effect of fixing $\gamma$ to 90 degrees, we also compute the
full Hessian matrix, including coupling to shear strains $\eta_4$ and $\eta_5$,
and plot  the lowest eigenvalue in Fig.~\ref{fig_energy}(d). 
It can be seen that the effect of the additional strain relaxation introduces a
narrow region of stability of the r-phase, between in-plane strains of 0.86$\%$ and 1.48$\%$. 
This window of 0.62$\%$ is about three times smaller than the T=0 width 
of 2$\%$ obtained by Pertsev et al.,\cite{Pertsev} 
the latter being an extrapolation to low temperature of a Landau theory
fit near the paraelectric-ferroelectric transition temperatures above 700 K.

\subsection{Lattice instabilities of the centrosymmetric reference structure}
\begin{figure}
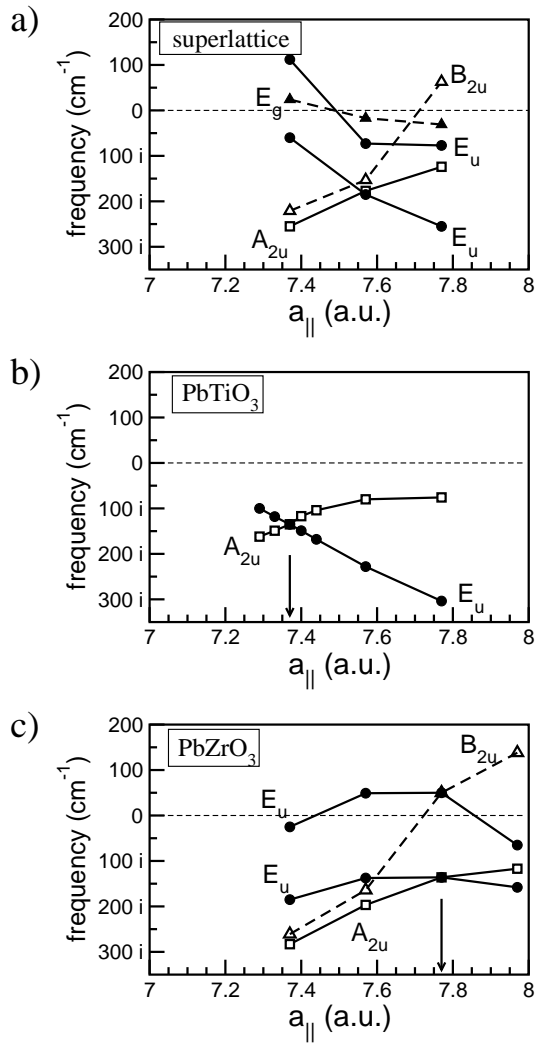

\vbox{
\centerline{\epsfig{figure=Fig4a.eps,width=7cm}}
\vspace{0.2cm}
\centerline{\epsfig{figure=Fig4b.eps,width=7cm}}
\vspace{0.2cm}
\centerline{\epsfig{figure=Fig4c.eps,width=7cm}}
}
\caption{Frequencies of the unstable zone-center phonons of the
centrosymmetric reference structure as a function of in-plane lattice constant
for (a) the [001]-(PbTiO$_3$)$_1$(PbZrO$_3$)$_1$ superlattice, (b) pure PT,
and (c) pure PZ. All modes are labeled
by their symmetry; solid (open) symbols indicate modes with displacements
perpendicular (parallel) to the c axis. Solid and dashed lines connect
the plotted points as a guide to the eye, indicating polar and nonpolar
character of the modes, respectively. Vertical arrows indicate the lattice
parameter of cubic PT and PZ.}
\label{fig_refmodes}
\end{figure}

Next, we investigate the relation of the ground state structure
to the zone-center phonon modes in the centrosymmetric reference structure. 
In the unstrained superlattice ($a_\parallel$=7.57 a.u.), 
there are two unstable, nearly degenerate, polar zone-center modes, 
an $E_u$ mode with displacements confined in the PT layer, 
and an $A_{2u}$ mode confined in the PZ layer.\cite{Bungaro,IrrepSymbols}
In Fig.~\ref{fig_refmodes}
we see that these behave differently with increasing in-plane strain, the 
PZ-like mode being stabilized by increasing in-plane strain, while the PT-like
mode becomes increasingly unstable. This is consistent with the qualitative
picture discussed in the introduction, with the mode with polarization along
the normal being raised in frequency by increasing in-plane lattice constant,
while the in-plane mode softens.
The ground state structure reflects these opposing trends in the strength of the
two unstable modes. At
the smallest in-plane lattice constant, it is energetically favorable to
freeze in only the $A_{2u}$ mode, producing the c-phase. As the in-plane
lattice constant increases, the in-plane $E_u$ mode softens to the point that
it also contributes to the ground state, producing the r-phase.
The frequencies of the two modes cross at an in-plane lattice constant
roughly in the middle of the r-phase.
With further increase of the in-plane lattice constant,
the $A_{2u}$ mode stiffens
to the point that the $E_u$ mode is the only mode contributing to the
ground state, producing the aa-phase.
The analysis of the Hessian matrix in the ground state structures shows that the
freezing-in of the polar modes eliminates the nonpolar instabilities also seen
in Fig.~\ref{fig_refmodes}, most notably the $B_{2u}$ mode.

The behavior of pure PT is qualitatively similar. There are two unstable polar 
zone-center modes $A_{2u}$ and $E_u$, increasing and decreasing in frequency with
increasing lattice constant, respectively. The crossing of the two modes is
likewise very close to the c-aa phase boundary, where the r-phase appears when
shear strain relaxation is allowed.

For comparison with the unstable modes in the superlattice, in Fig.~\ref{fig_refmodes}(c)
we show the 
in-plane strain dependence of the zone-center unstable modes of pure PZ.
For the unstrained superlattice we have previously shown~\cite{Bungaro} that 
the atomic displacements 
of the lowest E$_u$ mode, mainly confined in the PT layer, are very close to the
displacements of the polar unstable mode of PT bulk. Similarly the atomic displacements
of the lowest A$_{2u}$ mode, mainly confined in the PZ layer, are very close to the
displacements of the polar unstable mode of PZ bulk.
By comparing the phonon frequencies in Fig. ~\ref{fig_refmodes}, 
we see that certain modes in the superlattice can be 
identified with
modes in pure strained PT and PZ.
The PT-like mode of the superlattice, the lowest E$_u$ mode, has a strain-dependent
frequency very similar to that of the 
lowest E$_u$ mode of pure PT. 
The PZ-like mode of the superlattice, A$_{2u}$ mode, compares well with the 
A$_{2u}$ mode of pure PZ.
The B$_{2u}$ mode in the superlattice is a PZ-like mode with O-displacements mainly 
confined in the
PZ layer and its strain-dependent frequency closely corresponds to that of the B$_{2u}$ 
mode of pure PZ.
%
%

\subsection{Static dielectric tensor}
\begin{figure*} [!]
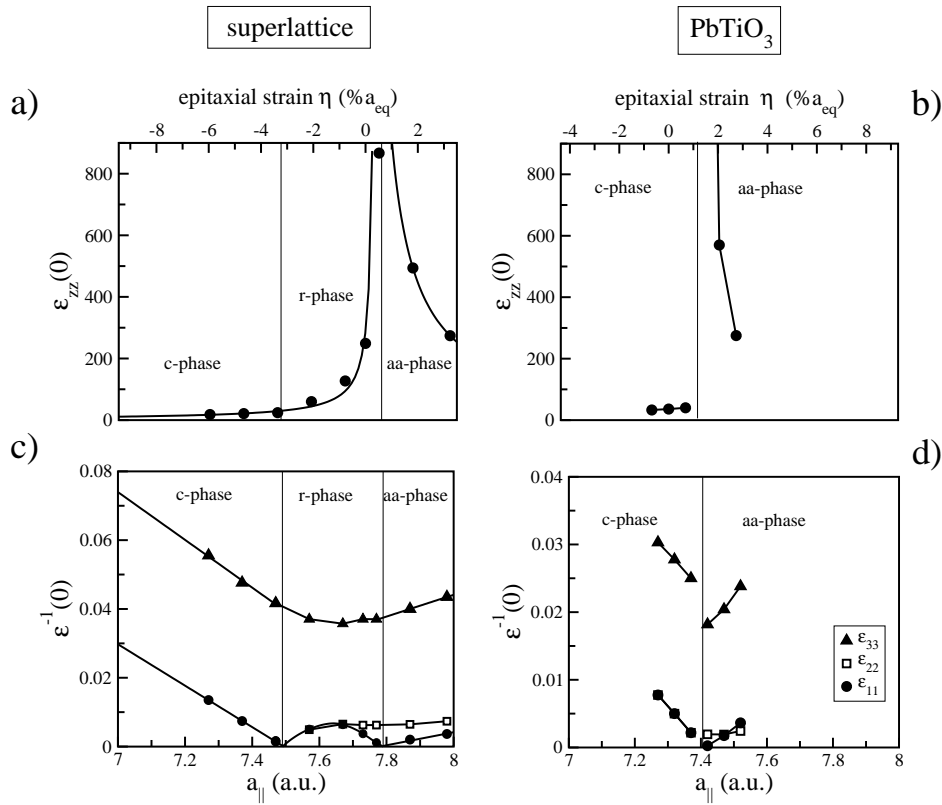

\vbox{
\centerline{
\hbox{
\epsfig{figure=Fig5a.eps,width=6cm}
\hspace{0.2cm}
\epsfig{figure=Fig5b.eps,width=6cm}
}
}
\centerline{
\hbox{
\epsfig{figure=Fig5c.eps,width=6cm}
\hspace{0.2cm}
\epsfig{figure=Fig5d.eps,width=6cm}
}
}
}
\caption{The phonon contribution to the static dielectric tensor 
component $\epsilon_{zz}$
as a function of the in-plane lattice constant
for (a) the [001]-(PbTiO$_3$)$_1$(PbZrO$_3$)$_1$ superlattice, and (b) pure PT.
Inverses of the eigenvalues of the phonon contribution to the dielectric tensor
as a function of the in-plane lattice constant
for (c) the [001]-(PbTiO$_3$)$_1$(PbZrO$_3$)$_1$ superlattice, and (d) pure PT.
The lines are a guide to the eye.
}
\label{fig_dielectric}
\end{figure*}

The phonon contribution to the static dielectric tensor 
computed as a function of epitaxial strain is presented for the
superlattice in Fig.~\ref{fig_dielectric}(a) and (c).
%
%
In (a), we plot the component $\epsilon_{zz}$, as for thin
films this is relevant to the dielectric screening of 
an electric field applied across the film.
It can be seen that this component diverges on both sides
of the r-aa phase boundary, and is large in a substantial
fraction of the r-phase.
In (c), we plot the inverses of the three eigenvalues of the 
dielectric tensor; thus a value approaching zero corresponds
to a divergence of $\epsilon$. 
In the c and aa phases, all three eigenvectors are uniquely
determined by symmetry. In the c-phase
$\epsilon_{3}=\epsilon_{zz}$ is the ``collinear" response, 
i.e. for a field applied along the 
direction of the spontaneous polarization, and 
$\epsilon_1=\epsilon_2=\epsilon_{xx}=\epsilon_{yy}$.
In the aa-phase, $\epsilon_{3}$ is the collinear component, along $[110]$,
$\epsilon_{2}$ is the in-plane component perpendicular to the 
polarization, and $\epsilon_{1}=\epsilon_{zz}$ is along the direction normal to the 
layers.
Two eigenvalues of the dielectric tensor ($\epsilon_1=\epsilon_2$) diverge at the 
continuous c-r transition, 
and one ($\epsilon_1$) at the continuous r-aa transition, through there the second noncollinear
eigenvalue $\epsilon_2$ is also fairly large.

In PT (Fig.~\ref{fig_dielectric}(b) and (d)), 
the behavior of the dielectric constant in the c and aa phases is rather
similar to that in the superlattice, with a near-divergence in $\epsilon_{zz}$
as the boundary of the aa-phase is approached. However, comparison of 
Fig. \ref{fig_dielectric}(a) and (b) show that the enhancement of the r-phase
in the superlattice gives a large $\epsilon_{zz}$ over a much larger range
of epitaxial strain.

\section{Discussion}

The most striking features of the phase diagram of the epitaxially constrained 
[001]-(PbTiO$_3$)$_1$(PbZrO$_3$)$_1$ superlattice are the width of the r-phase,
the separation of the minimum energy in-plane lattice constants of the c and aa-phases, 
and the linearity of the
energy with in-plane strain in the r-phase. These have significant implications for 
the coherence
of lattice-mismatched thin films and for the dielectric response.

From Fig.~\ref{fig_energy}(a), it can be seen that the elastic energy of the 
superlattice as a function of 
in-plane lattice constant is not parabolic. Rather, as the in-plane lattice
constant decreases from 7.73 a.u., the energy rises linearly with a small slope to the
minimum of the c-phase at 7.50 a.u., 
behaving parabolically only for smaller $a_\parallel$. This makes
the film compliant to a broad range of substrates; the estimated
critical thickness of fully coherent films is close to constant in this range. 
For tetragonal PbTiO$_3$, in comparison, within each phase (c or aa), the energy function
is parabolic, so the critical thickness behaves as usual, increasing inversely to the deviation
of the in-plane lattice constant from the equilibrium value for that phase.

Based on the computed energies for the various phases of the superlattice 
in Fig.~\ref{fig_energy}(a), one can predict the
occurrence of structural phase transitions with increasing thickness, as the system changes
from fully coherent to partially relaxed and finally to fully relaxed. For systems under
compressive stress from the substrate, there should be a transition from the c phase to the
r phase with increasing thickness, while for systems under tensile stress, a transition
from the aa phase to the r phase is expected. For pure PT, systems under tensile stress 
are predicted
to undergo a transition with increasing thickness from the aa-phase through a narrow r-phase 
into the c-phase. 

The present analysis identifies
the most stable superlattice structure with the 
ten-atom cell shown in Fig.~\ref{fig_supercell}. The possibility remains
that there may be instabilities to structural distortions that double or further expand 
the unit cell,
in the plane, along the stacking direction, or both. 
This may be relevant for making comparisons with experimental observations.

Certain components of the dielectric tensor diverge 
as the phase boundaries are approached. 
While the polarization induced by an electric field along the direction of the 
spontaneous polarization
is rather modest, the noncollinear responses are of great interest. 
This polarization rotation mechanism, through the coupling of polarization and strain, has
been invoked to explain the large piezoelectric response in single-crystal relaxors
\cite{FuCohen,Shrout}, and to explain the large piezoelectric response computed
for ordered PZT (with the same structure as the superlattice considered here) in 
Ref.~\onlinecite{KrakauerWu}. 
The r-phase, with its continuously variable direction for the spontaneous polarization,
is expected to have particularly large responses. 
In particular, the $zz$ component of $\epsilon$, the normal polarization induced by a 
field along the normal, will be enhanced by a noncollinear contribution throughout the r-phase.

Finally, we consider the origin of the different behavior of the two systems.
First, we discuss the difference in equilibrium lattice constant of the 
c-phase and the aa-phase.
In a simple perovskite, such as pure PT, the difference arises from coupling
between the strain and the unstable polar mode. PT has an unusually strong strain
coupling, leading to a $c/a$ of about 1.06 in the tetragonal ground state; the c-phase and aa-phase
equilibrium in-plane lattice constants are 7.32 a.u. and 7.45 a.u., 
respectively.
Analogous information about the behavior of pure PZ can be 
obtained from the published parameterization of first-principles results\cite{KingSmith} with
the c-phase and aa-phase
equilibrium in-plane lattice constants being 7.71 a.u. and 7.83 a.u., respectively.
A simple elastic-energy model for the superlattice can be constructed as the sum of 
the elastic energies of the two layers. For each constituent, the elastic constants 
of the c and aa phases turn out to be similar, with the values for PZ about 70\% larger 
than those for PT.
In the superlattice, the first-principles equilibrium in-plane lattice constant of the 
c-phase is 0.07 a.u. less than that of the elastic-energy model, and in the aa-phase it 
is 0.05 a.u. greater,
leading to a separation of the first-principles lattice constants of the two phases about
twice as large as in the model. This deviation from a simple elastic energy model 
might be due to the thinness of the individual
layers and/or to the importance of inter-layer interactions beyond lattice matching.

The stability of the r-phase is favored by in-plane and perpendicular
unstable polar modes having both a similar strength and positive
anharmonic coupling.
In a simple perovskite, such as pure PT, tetragonal strain splits the three-fold degenerate mode
to give a dominant unstable $A_{2u}$ mode for $c/a$ less than 1, and a dominant
in-plane $E_{u}$ mode for $c/a$ greater than 1. 
If the ground state is obtained
by freezing in only the dominant unstable mode, we expect a discontinuous change in character
at the in-plane lattice constant corresponding to the lowest-energy cubic structure,
as seen in PT.
Even if the strengths of the two modes are similar, the fourth order anharmonic term
coupling them must be large enough to favor this combination over the higher-symmetry
structures (z, as in the c-phase, or along [110]
as in the aa-phase). Its magnitude can vary greatly from material to material, being
enhanced by coupling to particular types of strain, especially shear strain.

In the superlattice, it seems that separating the two components of the
polarization in two spatially distinct regions helps to stabilize the r-phase structure.
Even if the strengths of the unstable polar modes in the two layer are different, 
a low-energy structure can be obtained by
combining the perpendicular mode with polarization along z in one layer 
(e.g. the PZ layer) with a mode of different character, with polarization
along [110], in the other layer (e.g. the PT layer). This idealized picture 
will be somewhat modified by electrostatic considerations, which will tend to 
minimize the divergence of {\bf P} by polarizing the PT layer along the z direction.

The usefulness of this simple picture lies in its applicability to
longer period PT/PZ superlattices; such systems are also of interest as they
are accessible to experimental investigation.

\section{Conclusions}
The phase diagrams for the [001]-(PbTiO$_3$)$_1$(PbZrO$_3$)$_1$ superlattice 
and pure PT with varying in-plane strain on a cubic substrate have been
obtained from first-principles density-functional total-energy calculations and
a stability analysis involving the computation of zone center phonons and their 
coupling to strain.
At the smallest $a_\parallel$, both systems are tetragonal with pseudo-perovskite 
$c/a$ greater than 1 and ${\bf P}$ along [001] (the c-phase), and
at the highest $a_\parallel$, both are pseudo-tetragonal orthorhombic with 
pseudo-perovskite $c/a$ less than 1 and
${\bf P}$ along [110] (the aa-phase). 
However, at intermediate in-plane strain, the transitional behavior is quite different.
In the superlattice,
${\bf P}$ rotates away from the normal to give a monoclinic structure (the r-phase),
and at a higher value of $a_\parallel$, completes the rotation into the plane of
the layers. In the monoclinic phase,
the dielectric and piezoelectric response is high, the sensitivity
to applied fields and stresses resulting from the ease with which the polarization
can be changed through rotation.
In contrast, in pure PT changing epitaxial strain results in a discontinuous 
transition from ${\bf P}$ along [001] to ${\bf P}$ in plane if only tetragonal strain
is allowed, with full coupling to shear strain opening a narrow range of r-phase
stability. While the characteristic divergence of certain components of the dielectric
tensor is seen as the phase boundaries are approached, there is no significant contribution
from polarization rotation.
For both systems, we relate the observed ground states to the instabilities of a relaxed
centrosymmetric structure computed via DFPT. 
From these results we conclude that in the superlattice,
the presence of two unstable modes confined in different layers,
of different symmetry and comparable strength 
significantly enhances the window for 
the monoclinic structure, which suggests new approaches for designing artificially
structured materials with high dielectric and piezoelectric response.

\begin{acknowledgments}
We thank D. Vanderbilt, J. Neaton, O. Dieguez, A. Antons, S. Tinte,
M. H. Cohen and H. Krakauer for valuable discussions.
This work was supported by ONR N00014-00-1-0261. The work of
K.M.R. was performed in part at LMCP, Universit\'e de Paris VI and Materials Department,
University of California at Santa Barbara. The
majority of the computations were performed at the Department of Defense High 
Performance Computing Centers NAVO and ERDC.
\end{acknowledgments}


\begin{references}
\bibitem{IrrepSymbols} We have classified the phonon modes according to the 
irreducible representations of P4/mmm, using the Mulliken symbols 
E$_u$, A$_{2u}$, B$_{2u}$, and E$_g$ that correspond to the modes
$\Gamma_5^\prime$, $\Gamma_1^\prime$, $\Gamma_3^\prime$, and $\Gamma_5$, 
respectively, in the nomenclature we used in Ref.~\onlinecite{Bungaro}.
\end{references}
\end{document}